\documentclass[tikz,11pt,english,a4paper]{article}
\usepackage{float}
\usepackage{jcappub}
\usepackage[utf8]{inputenc}

\usepackage{bm}
\usepackage{algorithm}
\usepackage{nicefrac}
\usepackage{tabto}
\usepackage[linewidth=1pt]{mdframed}
\usepackage{setspace}
\usepackage{multirow}

\usepackage[noend]{algpseudocode}
\makeatletter
\def\BState{\State\hskip-\ALG@thistlm}
\makeatother

\usepackage{fontawesome5}
\usepackage{hyperref}
\makeatletter
\newcommand{\github}[1]{%
   \href{#1}{\faGithub}%
}
\makeatother

\usepackage{lmodern}

\newcommand{\tf}{TensorFlow}

\usepackage{siunitx}
\DeclareSIUnit \parsec {pc}

\usepackage[T1]{fontenc}

\usepackage{soul}

\usepackage[edges]{forest}
\definecolor{folderbg}{RGB}{124,166,198}
\definecolor{folderborder}{RGB}{110,144,169}
\newlength\Size
\tikzset{%
  folder/.pic={%
    \filldraw [draw=folderborder, top color=folderbg!50, bottom color=folderbg] (-1.05*\Size,0.2\Size+5pt) rectangle ++(.75*\Size,-0.2\Size-5pt);
    \filldraw [draw=folderborder, top color=folderbg!50, bottom color=folderbg] (-1.15*\Size,-\Size) rectangle (1.15*\Size,\Size);
  },
  file/.pic={%
    \filldraw [draw=folderborder, top color=folderbg!5, bottom color=folderbg!10] (-\Size,.4*\Size+5pt) coordinate (a) |- (\Size,-1.2*\Size) coordinate (b) -- ++(0,1.6*\Size) coordinate (c) -- ++(-5pt,5pt) coordinate (d) -- cycle (d) |- (c) ;
  },
}
\forestset{%
  declare autowrapped toks={pic me}{},
  declare boolean register={pic root},
  pic root=0,
  pic dir tree/.style={%
    for tree={%
      folder,
      font=\ttfamily,
      grow'=0,
    },
    before typesetting nodes={%
      for tree={%
        edge label+/.option={pic me},
      },
      if pic root={
        tikz+={
          \pic at ([xshift=\Size].west) {folder};
        },
        align={l}
      }{},
    },
  },
  pic me set/.code n args=2{%
    \forestset{%
      #1/.style={%
        inner xsep=2\Size,
        pic me={pic {#2}},
      }
    }
  },
  pic me set={directory}{folder},
  pic me set={file}{file},
}

\usepackage{newfloat}
\usepackage{caption}
\DeclareFloatingEnvironment[fileext=frm,placement={!t},name=Algorithm]{pseudoenv}
\DeclareCaptionLabelSeparator{normal-colon}{\normalfont : } 
\newenvironment{pseudo}[2]{
    \gdef\tempcaption{#1}
    \gdef\templabel{#2}
    \begin{pseudoenv}[tb]
    \begin{mdframed}[roundcorner=10pt, middlelinewidth=1pt]
    \begin{center}
    \begin{tabular}{l|l}}
    {\end{tabular}
    \end{center}
    \end{mdframed}
    \caption{\tempcaption}
    \label{\templabel}
    \end{pseudoenv}
    }

\usepackage{cleveref}[2012/02/15]
\crefformat{footnote}{#2\footnotemark[#1]#3}

\begin{document}


\title{Fast and efficient nested sampling with BEST}

\author[a]{Andreas Nygaard}

\affiliation[a]{Department of Astrophysics, University of Zurich,
 8057 Zurich, Switzerland}

\emailAdd{andreas.hansen@uzh.ch}

\abstract{Nested sampling is widely used for Bayesian evidence computation, but its intrinsically sequential structure limits how efficiently it can exploit modern vectorised likelihoods and emulators. We present a new nested-sampling implementation in \textsc{best}, written entirely in TensorFlow and designed for efficient XLA compilation on both CPUs and GPUs. The sampler combines clustering and slice sampling with the possibility of updating several live points simultaneously. Since batching breaks the strict ordering of conventional nested sampling, we introduce sorting and history-based corrections to reduce the resulting bias in the evidence estimate.

We test the sampler on Gaussian, Rosenbrock, and multimodal likelihoods and compare its performance with JAXNS and UltraNest. The results show that accurate evidence estimates can be retained for moderate batch sizes, with $m/N_{\rm live}\lesssim 0.1$ providing a useful practical regime. Finally, using a 27-dimensional cosmological likelihood emulator, we show that batched live-point updates can substantially reduce the wall-clock time while remaining consistent with sequential sampling within the reported uncertainties. The new implementation therefore extends \textsc{best} with an efficient nested-sampling method tailored to fast, vectorised likelihoods and emulator-based inference.
}

\maketitle

\section{Introduction}\label{sec:introduction}

Nested sampling has since its introduction in Ref.~\cite{Skilling:2006gxv} been a vital part of model selection within cosmology. Determining the prior volume accurately is very important for estimating the Bayesian evidence, and nested sampling provides an efficient way of integrating the entire multidimensional likelihood over the prior volume, whereas standard Markov chain Monte Carlo (MCMC) methods primarily sample the posterior and cannot by themselves be used to evaluate the evidence~\cite{2011MNRAS.414.1418K}. The general idea behind nested sampling is to reformulate the problem as a one-dimensional integral instead:
\begin{equation}\label{eq:evidence}
	\mathcal{Z} = \int_0^1 \mathcal{L}(X) {\rm d}X\,,
\end{equation}
where the evidence, $\mathcal{Z}$, is computed as an integral over the likelihood function, $\mathcal{L}$, w.r.t. the prior volume, $X$. Computing the evidence through nested sampling thus requires sampling $\{\mathcal{L}_i, X_i\}$ pairs where $\mathcal{L}_i$ is monotonically increasing for decreasing $X_i$. This is done by continuously updating a set of $N_{\rm live}$ live points by removing the worst point and sampling a new point with a better likelihood. Each iteration then represents a nested slice of the distribution and from the total set of nested slices, the integral can be computed.

Many different implementations of this are publicly available, such as \textsc{MultiNest}~\cite{Feroz:2008xx}, \textsc{PolyChord}~\cite{2015MNRAS.453.4384H}, \texttt{dynesty}~\cite{Speagle:2019ivv}, UltraNest~\cite{Buchner:2021cql}, and JAXNS~\cite{2020arXiv201215286A}, and the different implementations are optimised for different problems, e.g., \textsc{MultiNest} is optimised for multimodal distributions in low dimensions due to its clustering algorithm and ellipsoidal sampler, while \texttt{dynesty} uses \emph{dynamic nested sampling}~\cite{Higson:2018cwj} to vary the number of live points depending on how important the current iteration is to the result.

In the past decade, emulation has grown increasingly popular with the advances of machine learning, and within cosmology, numerous emulators and emulation frameworks exist~\cite{Nygaard:2022wri,SpurioMancini:2021ppk,Janken:2025wlq,Gammal:2022eob,Gunther:2025xrq,Gunther:2022pto,Cohen:2026iij}. These emulators can replace conventional forward models for inference purposes, and this has been extensively demonstrated both for Bayesian parameter inference~\cite{Nygaard:2026fgl,Nygaard:2022wri,SpurioMancini:2021ppk,ManciniSpurio:2021jvx,Lynch:2024gmp,Janken:2025wlq} and Bayesian evidence computation~\cite{Sorensen:2025ywu,SpurioMancini:2021ppk,Cohen:2026iij,Lovick:2025wdj,Carrion:2024itc,Piras:2023aub}. A highly useful property of such emulators is their efficient vectorisation which allows parallelisation at little-to-no extra computational cost. For MCMC sampling, this can greatly improve the efficiency by allowing multiple chains to run vectorised on the same GPU~\cite{Nygaard:2026fgl}.

Nested samplers have conventionally been limited by the cost of the forward model and subsequent likelihood computation, but with emulators, the computational overhead of the nested sampler starts to matter very much to the total wall clock time. In order to fully utilise the immense speed of the emulators, nested samplers tailored to the same frameworks are important. An example of this is JAXNS~\cite{2020arXiv201215286A} which is purely written in Google's JAX framework~\cite{jax2018github}. Many emulators are using this framework due to its versatility and hardware exploitation. JAXNS is able to achieve runtimes much lower than its conventional counterparts, but it is still limited by the algorithmic properties of nested sampling and only utilises vectorisation to some extent.

In this paper, we present a new nested sampler that similarly to JAXNS is implemented in an efficient framework, namely Google's TensorFlow framework~\cite{tensorflow}. Also like JAXNS, it uses efficient hardware exploitation through the \emph{Accelerated Linear Algebra} (XLA) compiler in order to fuse operations efficiently into a compiled computational graph, thus leading to huge speed-ups for continuous usage. Unlike JAXNS (and most other nested samplers), it does not adhere to the strict algorithmic sequentiality that ensures a correct monotonically increasing behaviour of $\mathcal{L}_i$ for decreasing $X_i$. Instead it takes advantage of the vectorisation properties of emulators (and other TensorFlow functions) to update $m$ live points simultaneously. Depending on $m/N_{\rm live}$, the resulting evidence might be biased due to a possible wrong ordering, but this is then corrected for through sorting and replacement of invalid points. The simultaneous replacement of several live points has previously been considered in Refs.~\cite{BURKOFF2012878,10.1063/1.4903717}. In particular, Ref.~\cite{10.1063/1.4903717} removes $r$ live points simultaneously and uses the best likelihood among these as a common constraint for all replacement points. The corresponding prior-volume shrinkage distribution is modified consistently for the removal of $r$ points. For a fixed number of live points, however, its variance increases with $r$, and Ref.~\cite{10.1063/1.4903717} shows that approximately $N_r\simeq\sqrt{r}N_1$ live points are required to retain the precision of a sequential calculation with $N_1$ live points. More recently, this approach has been developed into a fully vectorised, GPU-accelerated nested sampler in Ref.~\cite{2026arXiv260123252Y}, where batched deletion is combined with vectorised constrained slice sampling. Like Ref.~\cite{10.1063/1.4903717}, all replacement points are generated using the best likelihood among the removed points as a common constraint. Ref.~\cite{2026arXiv260123252Y}, however, assigns the removed points individual prior-volume contractions by unrolling the batch into sequential single-death events for the evidence calculation. Our approach is different in that we additionally attempt to retain the individual likelihood constraints that would arise during sequential live-point replacement. This introduces the ordering problem described above, but, if corrected sufficiently accurately, allows the simultaneous updates to retain a precision comparable to sequential nested sampling with the same number of live points. This is particularly advantageous for vectorised likelihood functions, for which several likelihood evaluations can be performed simultaneously at comparatively little additional computational cost.

The structure of the paper is as follows: In Section~\ref{sec:implementation}, we sketch the overall implementation, in Section~\ref{sec:batch}, we elaborate on the batched updates of live points and the corrections implemented to adjust for it, in Section~\ref{sec:comparison}, we compare the nested sampler with JAXNS and UltraNest on a set of standard problems, and in Section~\ref{sec:cosmo}, we use the nested sampler with a cosmological likelihood emulator trained with the CLiENT framework~\cite{Janken:2025wlq}. Finally, we draw our conclusions in Section~\ref{sec:conclusion}.

\section{Implementation}\label{sec:implementation}
The nested sampler is implemented in the \tf{}-based package \textsc{best}\footnote{Installable from GitHub: \url{https://github.com/AndreasNygaard/best-inference}, or from PyPI using \texttt{pip install best-inference}} and is optimised for both CPU and GPU usage. The operations are written to allow for them to be efficiently compiled using XLA fusion.

\subsection{Clustering of live points}
An important part of a nested sampler is a way of identifying different modes in the distribution. If this is not accounted for and the distribution features multiple disjoint modes, new proposals may become almost impossible to generate -- especially in high dimensionality. As proposed in the Refs.~\cite{Feroz:2007kg,Feroz:2008xx}, it is beneficial to divide the points into clusters from which proposals can be generated. Following the same procedure, we use the $k$-means clustering algorithm~\cite{lloyd1982least} with $k=2$ to iteratively split the live points into several clusters. As pointed out in Ref.~\cite{2020arXiv201215286A}, the XLA compiled parts of the implementation must be static-memory, which means that all array shapes and data types must be known at compile time. It is therefore not straightforward to use the same clustering approach as \textsc{MultiNest} and \textsc{PolyChord} where the number of clusters can be arbitrary depending on the distribution of live points in a specific iteration. Instead, we employ a method similar to Ref.~\cite{2020arXiv201215286A}, where a binary tree of clusters is initialised with a user-specified maximum depth, where a depth of 0 corresponds to just 1 cluster and a depth of $d$ corresponds to a maximum of $2^d$ clusters with $2^{d+1}-1$ total clusters in the tree. All $k$-means splittings are then performed throughout the binary tree and the volumes of the individual clusters are computed. The tree is then pruned based on the volumes from the bottom up, where two child-clusters are favored over their parent only if their combined volume is more than a user-specified percentage smaller than the volume of the parent-cluster and if both contain more than the minimum number of points for a cluster (also user-specified). The rejected clusters are then masked away, thus keeping the array shapes of later computations invariant and allowing for the clustering algorithm to discover a different number of clusters in later iterations.

In order to compute the volume, the covariance must be computed for each cluster. This covariance can then later be reused for slice sampling, but it is not guaranteed to be numerically stable due to floating-point errors, rank deficiency (clusters with fewer points than the dimensionality), or the fact that covariance matrices are only guaranteed to be positive-semidefinite (PSD) whereas the standard Cholesky routine used here requires a positive-definite (PD) matrix~\cite{benoit1924cholesky}. We therefore slightly adjust problematic covariance matrices by eigenvalue-clipping to find the nearest PSD matrices and regularise them to obtain PD covariance matrices and their Cholesky decompositions.

Even though the clustering itself is quite fast and effective, it is strictly not needed in every iteration, given that the set of live points only changes slightly. The sole purpose of the clustering is to obtain proposals more efficiently, so keeping stationary cluster covariance for several iterations does not significantly impact the efficiency of generating new live points. The new live points need to be correctly assigned to the existing clusters though, in order for an appropriate covariance to be used for slice sampling if a new point is chosen as a seed for the slice sampler (see Section~\ref{sec:slice}) before reclustering. This assignment is done by identifying the accepted cluster whose center is closest to the new point.

\subsection{Slice sampling}\label{sec:slice}
Once the live points have been divided into appropriate clusters, a new proposal point can be generated in multiple ways. Refs.~\cite{Feroz:2007kg,Feroz:2008xx} use ellipsoidal sampling where a proposal is drawn uniformly from the union of ellipsoids enclosing the clusters. This is very advantageous in low dimensions, but as it relies on rejection sampling within the enlarged ellipsoids, the \emph{curse of dimensionality} makes this sampling strategy increasingly inefficient at high dimensionality~\cite{2015MNRAS.453.4384H}.

Slice sampling, introduced in Ref.~\cite{2000physics...9028N}, provides an alternative that scales much better with dimensionality. A new proposal is generated by choosing a random live point (seed point) and a random direction in which to bisect the probability landscape. An initial bracket along this one-dimensional slice is constructed and expanded until its endpoints lie outside the $\mathcal{L}>\mathcal{L}_{\rm worst}$ region along the chosen direction. A uniform point in this bracket is then drawn and if it is valid, it is kept, and if not, the bracket is shrunk to the position of the drawn point. This repeats until a valid sample has been drawn. Doing just one slice-sampling step, however, ends up correlating the new point with the seed point, so to avoid this, the new valid point is used as a seed point in a new slice sampling with a new random direction. This should be repeated a sufficiently large number of times, usually $n\times d$ where $d$ is the dimensionality and $n$ is a small positive integer of order 1 to order 10, in order to reduce the correlation with the seed point~\cite{2015MNRAS.453.4384H}. 

When all slice samplings are done, the resulting point is approximately distributed according to the constrained prior, with reduced correlation with the seed point. The live point with $\mathcal{L}_{\rm worst}$ is then replaced with the new point in the set of live points.

The clustering of live points comes in handy when the distribution contains different modes whose individual orientations are different. Sampling effectively from the modes thus requires that the covariance of each mode is taken into account. When choosing the random live point from which to start the slice sampling, the random direction to sample along can be chosen isotropically and then transformed using the covariance of the cluster to which the particular live point belongs. In the case of strong correlations, this can help sample uniformly from the cluster in much fewer slice steps. Given enough slice steps, the slice sampler will perform well regardless of whether or not clustering is performed, but for multimodal or very complex likelihood landscapes, clustering can reduce the number of required likelihood evaluations. This is further explored in Section~\ref{sec:cluster}.

\subsection{Significance of clustering}\label{sec:cluster}

To assess how important clustering is, we test the nested sampler on two multimodal likelihoods as well as a more complex toy model. The first one (Test 1) is an array of 49 two-dimensional Gaussians in a $[-20,20]^2$ uniform prior box:
\begin{equation}\label{eq:array}
	\log{\mathcal{L}}(\vec{x}) = \log{\left[\sum_{i=-3}^3\sum_{j=-3}^3\exp{\left(-\frac{1}{2}\left[(x_1 - 4i)^2 + (x_2 - 4j)^2\right]\right)}\right]}\,,
\end{equation} 
the second one (Test 2) is comprised of two 10-dimensional Gaussians of different heights with a separation much larger than their widths in a $[-20,20]^{10}$ uniform prior box:
\begin{align}\label{eq:separation}
	\nonumber\log{\mathcal{L}}(\vec{x}) &= \log{\left[\frac{1}{10}{\rm e}^{\left(-\frac{1}{2}(\vec{x}-\vec{\mu}_1)^{\rm T} \Sigma_1^{-1} (\vec{x}-\vec{\mu}_1) \right)} + {\rm e}^{\left(-\frac{1}{2}(\vec{x}-\vec{\mu}_2)^{\rm T} \Sigma_2^{-1} (\vec{x}-\vec{\mu}_2)\right)}\right]}\,, \\
	\nonumber\mu_1 &= -\mu_2 = [-10,\ldots,-10]\,,\\
	\Sigma_1^{(2k,2k)} &= \Sigma_2^{(2k-1,2k-1)} = 1 \hspace{3em}({\rm for}\; k=1,\ldots,5)\,, \\
	\nonumber\Sigma_2^{(2k,2k)} &= \Sigma_1^{(2k-1,2k-1)} = 0.1 \hspace{2.22em}({\rm for}\; k=1,\ldots,5)\,, \\
	\nonumber\Sigma_1^{(i,j)} &= \Sigma_2^{(i,j)} = 0 \hspace{5.99em} ({\rm for}\; i\neq j) \, ,
\end{align}
while the third one (Test 3) is a two-dimensional Gaussian shell likelihood in a $[-10,10]^2$ uniform prior box:
\begin{equation}\label{eq:shell}
	\log{\mathcal{L}}(\vec{x}) = -\frac{1}{2}\left(\sqrt{x_1^2 + x_2^2}-5\right)^2\, .
\end{equation} 

For each of these tests, we run 50 nested samplings using 1,000 live points and record the means and standard deviations of the resulting evidences, wall clock times, and numbers of likelihood evaluations, both using clustering and omitting it. These tests are run exclusively on CPU (14-core Apple M4 Pro), since the simplicity of the models allows for much smaller run-times than one would obtain on a GPU with the same configurations. When using clustering, we restrict the clusters to only update every 100 iterations to keep down the most dominant cost of it, and we use a tree depth of $d=6$ (maximally 64 clusters) for Test 1, a tree depth of $d=1$ (maximally 2 clusters) for Test 2, and a tree depth of $d=4$ (maximally 16 clusters) for Test 3. The tree depths have been chosen corresponding to the smallest tree able to accurately accommodate the structure of each likelihood. The results are displayed in Table~\ref{tab:slice}.
\begin{table}[]
\centering
\begin{spacing}{1.2}
\begin{tabular}{lc|cccc}
                  &                & $\bm{\log{\mathcal{Z}}}$  & $\bm{\log{\mathcal{Z}_{\rm true}}}$ & \textbf{Wall clock time {[}s{]}} & \textbf{Evaluations} \\ \hline
\multirow{2}{*}{\textbf{Test 1}} &$\bm{(+)}$ & $-1.6470 \pm 0.0207$ & \multirow{2}{*}{$-1.6481$}                       & $1.027 \pm 0.013$                     & 102,881      \\
                                                &$\bm{(-)}$   & $-1.6492 \pm 0.0212$ &                                                                & $0.113 \pm 0.002$                     & 124,568       \\ \hline
\multirow{2}{*}{\textbf{Test 2}} &$\bm{(+)}$ & $-33.408 \pm 0.198$  & \multirow{2}{*}{$-33.360$}                        & $3.49 \pm 0.41$                         & 10,508,225   \\
                                                &$\bm{(-)}$  & $-33.272 \pm 0.221$  &                                                                  & $4.03 \pm 0.29$                         & 14,144,037   \\ \hline
\multirow{2}{*}{\textbf{Test 3}} &$\bm{(+)}$ & $-1.6261 \pm 0.0274$ & \multirow{2}{*}{$-1.6252$}                       & $0.273 \pm 0.005$                     & 96,925         \\
                                                &$\bm{(-)}$  & $-1.6322 \pm 0.0314$ &                                                                 & $0.048 \pm 0.002$                     & 134,944              

\end{tabular}
\end{spacing}
\caption{\textsl{Evidence results, wall clock times (on 14-core Apple M4 Pro CPU), and numbers of likelihood evaluations for the array of Gaussians (Test 1) described by Eq.~\eqref{eq:array}, the two very separated Gaussians (Test 2) described by Eq.~\eqref{eq:separation}, and the Gaussian shell (Test 3) described by Eq.~\eqref{eq:shell}. The results are shown both when including clustering ($+$) and when omitting it ($-$). Each configuration is based on 50 independent nested samplings of which the means and standard deviations make up the recorded results.}}
\label{tab:slice}
\end{table}
The table clearly shows that clustering is able to bring down the total number of likelihood evaluations, but the resulting evidence is not significantly affected. The results of Test 1 are almost identical and the most noticeable difference is in the wall clock times where including clustering slows down the sampling by a factor of $\sim \!9$. Test 1 is a two-dimensional likelihood, and even though clustering reduces the overall number of likelihood evaluations, these are not dominant compared to the added costs of keeping track of clusters, so for very fast likelihoods and low dimensions, clustering does not benefit the sampling. Test 2, on the other hand, shows a speed-up when using clustering, and this can be attributed to the very significant reduction in the number of likelihood evaluations as well as a simpler clustering tree. Increasing the tree depth of Test 2 to $d=6$ does not change the result or the number of evaluations, but the wall clock time increases drastically to $16.0\pm0.2$ seconds. Test 3 is also a two-dimensional likelihood, and the trend here is similar to that of Test 1, although clustering seems to yield a slightly better evidence result. This variation is, however, much less than the standard deviation between independent nested samplings. The use of clustering is thus primarily beneficial if the likelihood function is complex and costly enough or the tree depth is limited to only a few clusters.

\begin{figure}[tb]
	\centering
	\includegraphics[width=\textwidth]{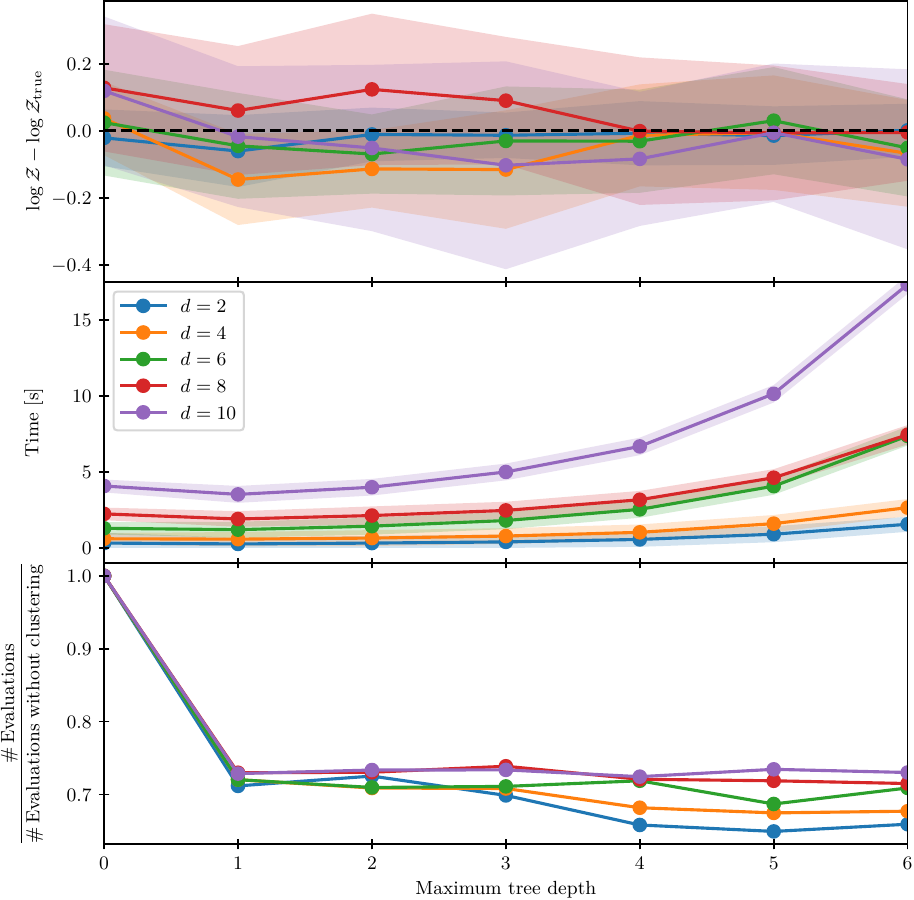}
	\caption{\textsl{Evidences (top), wall clock times (middle), and ratio between the number of likelihood evaluations for a given maximum tree depth and the number of evaluations when the tree depth is zero, i.e., without clustering (bottom), for the bimodal likelihood of Test 2 truncated to lower dimensions. All results are the means and standard deviations of 10 independent nested samplings using 1,000 live points. The samplings were performed on a 14-core Apple M4 Pro CPU.}}
	\label{fig:clustering}
\end{figure}

The dimensionality of the parameter space is worth considering when assessing the usefulness of clustering. For a more direct comparison, we reuse the bimodal likelihood from Test 2 but truncate the dimensionality to $d\in \{2,4,6,8,10\}$. We run 10 independent nested samplings for each configuration and the means and standard deviations of these are shown in Figure~\ref{fig:clustering}. The top panel shows the evidence as a function of the maximum tree depth for each dimensionality, and from this, the tree depth does not appear to affect the evidence result. The error on the evidence does, however, increase with dimensionality, but this is expected as all the samplings were performed with 1,000 live points, and increasing the number of live points for the 10-dimensional case does indeed decrease the error. The middle panel shows the time in seconds as a function of maximum tree depth, and, as one might expect, a larger tree depth takes longer to run. The total number of clusters in the tree scales exponentially with depth, so we should see a trend similar to this. The bottom panel shows the ratio between the number of likelihood evaluations for a given depth and the number of evaluations without clustering. Even though the absolute number of evaluations differs substantially between dimensionalities, the trend is the same: allowing for a minimum of two clusters (a depth of 1) brings down the number of evaluations by $\sim30\%$, and this does not change when increasing the depth due to two clusters accurately describing the distribution.

\section{Batching of live point updates}\label{sec:batch}
Nested sampling is fundamentally a sequential process in order to keep the correct ordering of the $\{\mathcal{L}_i, X_i\}$ pairs. Given a set of live points $S_i=\{x_1,...,x_N\}$ belonging to the $i^{\rm th}$ iteration and ordered from worst to best likelihood, replacing the worst live point, $x_1$, may result in a new sampled point, $\tilde{x}_1$, that is only slightly better and it could potentially be the next point to be removed, i.e. the ordered set could now be $S_{i+1}=\{\tilde{x}_1, x_2,...,x_N\}$. The very existence of $\tilde{x}_1$ in $S_{i+1}$ is however conditioned on $x_1$ being removed and used as a likelihood threshold when slice sampling in the previous iteration. If, however, a batch of $m$ live points were to be removed in the $i^{\rm th}$ iteration instead, all the points $\{x_1,...,x_m\}$ would be removed from $S_i$ and be replaced with the new sampled points $\{\tilde{x}_1,...,\tilde{x}_m\}$, which would result in adding $\tilde{x}_1$ to $S_{i+1}$ even though it is a worse point (given the same RNG seed as before) than the $m-1$ of the removed points of the previous iteration. This would break the ordering assumed by the conventional single-point evidence integration used here. Depending on the problem and the ratio $m/N$, this naïve approach can give rise to an extra source of error in the evidence result.

\subsection{Sorting and history correction}

A solution to the ordering problem could be to use the best likelihood in the batch of removed points as a likelihood threshold, i.e. $\mathcal{L}>\mathcal{L}_m$, for all slice samplings in the batch. This would completely remove the ordering problem since all new points would be better than all removed points. However, it introduces a new problem as well if the conventional sequential prior-volume assignment is retained (see Refs.~\cite{10.1063/1.4903717,2026arXiv260123252Y}). In the end, it would result in a set of $\{\mathcal{L}_i, X_i\}$ pairs where many likelihood values are too high since they were sampled with too strict of a constraint. From equation~\eqref{eq:evidence}, this results in an inflated evidence result. We shall refer to this approach as the \emph{best-of-worst} correction.

\begin{pseudo}{\textsl{Batch sorting approach when updating $m$ live points simultaneously. The algorithm sketches how to obtain the live points of the $(i+1)^{\rm th}$ iteration given the live points of the $i^{\rm th}$ iteration while correcting the ordering of the sampling.}}{alg:sort}
	\hspace{-1.4em}
	\begin{minipage}{0.57\textwidth}
		\vspace{0.5em}
		\begin{spacing}{1.2}
			\begin{algorithmic}
				\State $d$, $m$ \tabto{5em}$=$\hspace{0.23em} dimension, batch size
				\State $S_i$\tabto{5em}$=$\hspace{0.23em} $[x_1, x_2, ..., x_N]$\tabto{17.6em}$(N\times d)$
				\State $\mathcal{L}_i$\tabto{5em}$=$\hspace{0.23em} $[l_1, l_2, ..., l_N]$\tabto{19.1em}$(N)$
				\State $\bar{\mathcal{L}}$, $\bar{S}$ \tabto{5em}$=$\hspace{0.23em} \texttt{sorted}$(\mathcal{L}_i, S_i)$
				\State $S_W$\tabto{5em}$=$\hspace{0.23em} $\bar{S}[0:m]$ \tabto{17.6em}$(m\times d)$
				\State $W$\tabto{5em}$=$\hspace{0.23em} $\bar{\mathcal{L}}[0:m]$ \tabto{19.1em}$(m)$
				\State $S_{\rm random}$\tabto{5em}$=$\hspace{0.23em} \texttt{RNG.choose}$(m$ from $S_i)$
				\State $N$, $S_N$\tabto{5em}$=$\hspace{0.23em} \texttt{SliceSampling}$(S_{\rm random})$
				\vspace{1.5em}
				\State \textbf{def} \texttt{sort}$(arrays)$:
					\State \tabto{1.5em} \textrm{sorted} \tabto{5em}$=$\hspace{0.23em} \texttt{sort}$([arr$ \textbf{for} $arr$ \textbf{in} $arrays])$
					\State \tabto{1.5em} \textrm{reject} \tabto{5em}$=$\hspace{0.23em} $[arr[0:m]$ \textbf{for} \textit{arr} \textbf{in} $arrays]$
					\State \tabto{1.5em} \textrm{accept} \tabto{5em}$=$\hspace{0.23em} $[arr[m:2m]$ \textbf{for} \textit{arr} \textbf{in} $arrays]$
					\State \tabto{1.5em} \textbf{return} \textrm{accept}, \textrm{reject}
				\vspace{1.5em}
				\State $S_C$ \tabto{5em}$=$\hspace{0.23em} \texttt{concatenate}$([S_W, S_N])$
				\State $C$ \tabto{5em}$=$\hspace{0.23em} \texttt{concatenate}$([W, N])$
				\State $\mathcal{L}_{\rm acc}, S_{\rm acc}, \mathcal{L}_{\rm rej}, S_{\rm rej}$ \tabto{8.6em}$=$\hspace{0.23em} \texttt{sort}$(C, S_C)$
				\vspace{0em}
				\State $S_{i+1}$\tabto{5em}$=$\hspace{0.23em} \texttt{concatenate}$([S_{\rm acc}, \bar{S}[m:N]])$
				\State $\mathcal{L}_{i+1}$\tabto{5em}$=$\hspace{0.23em} \texttt{concatenate}$([\mathcal{L}_{\rm acc}, \bar{\mathcal{L}}[m:N]])$
				\State \texttt{dead\_points.append}$(S_{\rm rej})$
				\State \texttt{dead\_L.append}$(\mathcal{L}_{\rm rej})$
				\vspace{-1.2em}
			\end{algorithmic}
		\end{spacing}
	\end{minipage}
	&
	\hspace{0.07em}
	\begin{minipage}{0.40\textwidth}
		\vspace{2.53em}
		\begin{spacing}{1.2}
			\vspace{-1.2em}
			\textit{$N$ live points of $i^{\rm th}$ iteration}\\
			\textit{Likelihoods of live points}\\
			\textit{Sort according to $\mathcal{L}_i$}\\
			\textit{Worst $m$ points}\\
			\textit{Likelihoods of worst $m$ points}\\
			\textit{$m$ random points from $S_i$}\\
			\textit{New sampled likelihoods and points}\\

			\vspace{1.5em}
			\textit{Sort according to first array}\\
			\textit{First $m$ points are rejected}\\
			\text{Last $m$ points are accepted}\\

			\vspace{1.5em}
			\textit{Combined set of points to consider}\\
			\textit{Likelihoods of combined set}\\
			\textit{Get accepted and rejected points}\\
			\textit{Accepted points replace $S_W$ in $S_i$}\\
			\\
			\textit{Rejected points are dead points}\\
			\vspace{-1.2em}
		\end{spacing}
	\end{minipage}
\end{pseudo}

If insisting that each newly sampled point in the batch should be sampled with the corresponding likelihood thresholds, we can instead correct for the ordering error by combining the set of removed points and the set of newly sampled points into the set $C=\{x_1,...,x_m, \tilde{x}_1,...,\tilde{x}_m\}$ of size $2m$. If we order this set from worst to best likelihood, i.e. $\bar{C}=\{c^{(1)}, c^{(2)}, ..., c^{(2m)}\}$, the worst $m$ points, $\{c^{(1)}, ..., c^{(m)}\}$, can be removed as dead points while the best $m$ points, $\{c^{(m+1)}, ..., c^{(2m)}\}$, can be added to the set of live points. If there are any ordering problems to correct for, this means that some removed points will be moved back into the set of live points as \emph{revived points} due to worse points having been sampled by the slice sampler which in turn then are directly removed as \emph{instant deaths}. This, however, also introduces a new problem. If there are any revived points in the set of new points, $\{c^{(i)}\; |\; c^{(i)} \in \bar{C},\; i>m\}$, that means that some of the new points would have been sampled with the revived points' likelihood as a threshold even though the revived points ended up not being removed. Again this fixes the ordering but also leads to too high likelihood values in the $\{\mathcal{L}_i, X_i\}$ pairs which again produces an inflated evidence result. The number of affected pairs is, however, smaller in this case (depending on the ratio $m/N$), and so we expect the error to be as well. This approach, which we will refer to as \emph{batch sorting}, is sketched in Alg.~\ref{alg:sort}.

\begin{pseudo}{\textsl{Batch sorting approach with history correction. The algorithm sketches how to obtain the live points of the $(i+1)^{\rm th}$ iteration given the live points of the $i^{\rm th}$ iteration while correcting the ordering of the sampling as well as the inflation of likelihood values.}}{alg:history}
	\hspace{-1.4em}
	\begin{minipage}{0.60\textwidth}
		\vspace{0.5em}
		\begin{spacing}{1.2}
			\begin{algorithmic}
    				\State $d,m$ \tabto{5.5em}$=$ dimension, batch size
				\State $S_i$\tabto{5.5em}$=$\hspace{0.23em} $[x_1, x_2, ..., x_N]$ \tabto{18.8em}$(N\times d)$
				\State $\mathcal{L}_i$\tabto{5.5em}$=$\hspace{0.23em} $[l_1, l_2, ..., l_N]$ \tabto{20.2em}$(N)$
    				\State $\bar{\mathcal{L}}, \bar{S}$ \tabto{5.5em}$=$ \texttt{sorted}($\mathcal{L}_i, S_i$)
    				\State $S_W,\mathcal{L}_W$ \tabto{5.5em}$=$ $\bar{S}[0:m],\bar{\mathcal{L}}[0:m]$
				\State $P_W,H_W$ \tabto{5.5em}$=$ $[-1,\ldots,-1],[-1,\ldots,-1]$
    				\State $S_{\rm random}$ \tabto{5.5em}$=$ \texttt{RNG.choose}($m$ from $S_i$)
    				\State $N,S_N,H,S_H$ \tabto{6.5em}$=$ \texttt{SliceSampling}($S_{\rm random}$)
    				\State $P_N,H_N$ \tabto{5.5em}$=$ $\mathcal{L}_W,[0,\ldots,m-1]$
    				\State $S_C,\mathcal{L}_C$ \tabto{5.5em}$=$ \texttt{concatenate}($[S_W,S_N]$, $[\mathcal{L}_W,N]$)
    				\State $P_C,H_C$ \tabto{5.5em}$=$ \texttt{concatenate}($[P_W,P_N]$,$[H_W,H_N]$)
    				\State $\mathcal{L}_a,S_a,P_a,H_a,\mathcal{L}_r,S_r,P_r,H_r$
    				\State    \tabto{5.5em}$=$ \texttt{Alg1.sort}($\mathcal{L}_C,S_C,P_C,H_C$)
				\State
    				\For{$k=1,\ldots,K$}
        					\State $J_F$ \tabto{6em}$=$ $\{j:H_a[j]\geq0,\ P_a[j]\geq\mathcal{L}_a[0]\}$
        					\If{$J_F=\emptyset$}
            					\State \textbf{break}
        					\EndIf
        					\State $\bar{I}$ \tabto{6em}$=$ \texttt{sorted}($\mathcal{L}_r$ where $H_r\geq0$)
					\State $\tilde H,\tilde S_H$ \tabto{6em}$=$ $H[:,H_a[J_F]],S_H[:,H_a[J_F],:]$
					\State $\texttt{mask}$ \tabto{6em}$=$ $[\tilde H>\bar I]$
					\State $J_N[i]$ \tabto{6em}$=$ $\min\{j:\texttt{mask}[j,i]\}$
					\State $S_a[J_F[i]],\mathcal{L}_a[J_F[i]]$ \tabto{10em}$=$ $\tilde S_H[J_N[i],i],\tilde H[J_N[i],i]$
					\State $P_a[J_F[i]]$ \tabto{6em}$=$ $\bar I[i]$
					\State $S_C,\mathcal{L}_C$ \tabto{6em}$=$ \texttt{concatenate}($[S_r,S_a]$, $[\mathcal{L}_r,\mathcal{L}_a]$)
        					\State $P_C,H_C$ \tabto{6em}$=$ \texttt{concatenate}($[P_r,P_a]$, $[H_r,H_a]$)
        					\State $\mathcal{L}_a,S_a,P_a,H_a,\mathcal{L}_r,S_r,P_r,H_r$
        					\State    \tabto{6em}$=$ \texttt{Alg1.sort}($\mathcal{L}_C,S_C,P_C,H_C$)
    				\EndFor
				\State
    				\State $S_{i+1}$ \tabto{3.5em}$=$ \texttt{concatenate}($[S_a,\bar{S}[m:N]]$)
    				\State $\mathcal{L}_{i+1}$ \tabto{3.5em}$=$ \texttt{concatenate}($[\mathcal{L}_a, \bar{\mathcal{L}}[m:N]]$)
    				\State \texttt{dead\_points.append}($S_r$)
    				\State \texttt{dead\_L.append}($\mathcal{L}_r$)
				\vspace{-1.2em}
			\end{algorithmic}
		\end{spacing}
	\end{minipage}
	&
	\hspace{0.07em}
	\begin{minipage}{0.37\textwidth}
		\vspace{0.18em}
		\begin{spacing}{1.2}
			\textit{$N$ live points of $i^{\rm th}$ iteration}\\
			\textit{Likelihoods of live points}\\
			\textit{Sort according to $\mathcal{L}_i$}\\
			\textit{Worst $m$ points and $\mathcal{L}$s}\\
			\textit{Initialise worst parent arrays}\\
			\textit{$m$ random points from $S_i$}\\
			\textit{New points and shrink history}\\
			\textit{Initialise best parent arrays}\\
			\textit{Combine points and $\mathcal{L}$s}\\
			\textit{Combine parent $\mathcal{L}$s and indices}\\
			\textit{Use sort function from Alg.~\ref{alg:sort}}\\
			\\
			\\
			\textit{Do maximally $K$ iterations}\\
			\textit{Indices of faulty points}\\
			\textit{Halt if no faulty points}\\
			\\
			\textit{Sorted instant dead $\mathcal{L}$s}\\
			\textit{Shrink history for faulty points}\\
			\textit{Mask using new thresholds}\\
			\textit{Indices of new points in history}\\
			\textit{Points with relaxed thresholds}\\
			\textit{Update parent $\mathcal{L}$s}\\
			\textit{Combine all new points}\\
			\\
			\textit{Use sort function from Alg.~\ref{alg:sort}}\\
			
			\vspace{1.75em}
			\textit{Accepted replace $S_W$ in $S_i$}\\
			\\
			\textit{Rejected are dead}\\
			\vspace{-2.4em}
		\end{spacing}
	\end{minipage}
\end{pseudo}

In order to fully recover the sequential result, all points resulting in an ordering problem should be resampled with the correct threshold, but this will introduce dynamic batch sizes and is not well suited for the structure of this nested sampler. We therefore construct an approximation which requires storing samples from the shrinking phase of the slice sampler's very last iteration. This is exactly where the likelihood threshold is used to select the new point. If a new point is found to have been sampled with too high a threshold (from a revived point), the shrinking history is examined with a more relaxed likelihood threshold from one of the instant dead points. The earliest point in the shrinking process to satisfy $\mathcal{L}>\mathcal{L}_{\rm relaxed}$ will replace the new point in the live points. The relaxation only affects the last slice sampling iteration since it would be unfeasible to store the entire history of the slice sampling. This should, however, provide a good approximation since the first $n\times d - 1$ iterations serve to decouple the original seed point from the new point. For convex likelihood contours or situations where the relaxed threshold contour and the original threshold contour have roughly the same shape, this should not be a source of error since all points within the relaxed contour are reachable from some point within the original threshold via a single slice iteration. It is important, however, that the bracket from the expansion phase in the last slice sampling iteration contains the full relaxed likelihood interval, which can be ensured by always using the worst likelihood value of the batch in the expansion phase of the slice sampler. When choosing the relaxed likelihood threshold from the instant dead points, the lowest relaxed threshold is used for the new point with the lowest likelihood that is ``too good'', and so on. This results in a set of new points where some (the relaxed points) might now be worse than the best of the new dead points. To accommodate this, a final sorting of the combined set is done and the $m$ best points are added to the live points and the worst $m$ points are added to the dead points. This might introduce new ``too good'' points among the live points due to newly revived points resulting from the final sorting, though, and the solution to this is to continue this correction iteratively until all problematic points have been relaxed. The full iterative correction is sketched in Alg.~\ref{alg:history}, and we will refer to it as the \emph{history correction}. Even though the correction is rather cheap compared to the entire sampling, the default is to only use a single iteration of the history correction, since this is the leading order in most cases.

\subsection{Error estimation}
\begin{figure}[tb]
	\centering
	\includegraphics[width=\textwidth]{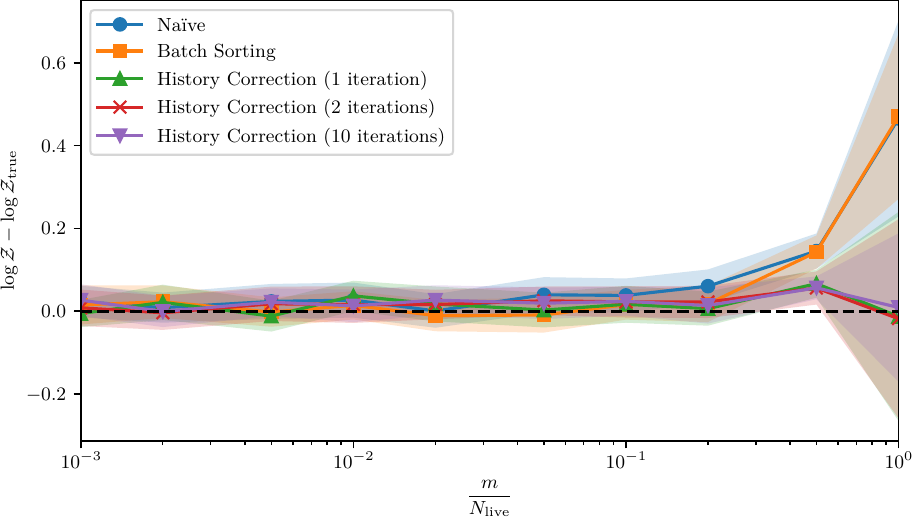}
	\caption{\textsl{Bayesian evidence results of the two-dimensional Gaussian in equation~\eqref{eq:gaussian} for different numbers, $m$, of simultaneously updated live points for $N_{\rm live}=10^3$ using different correction schemes. The lines represent the means of the evidences produced by 50 independent nested samplings with different random seeds and the bands represent the standard deviation of the same results. The dashed line represents the true value.}}
	\label{fig:cor1000}
\end{figure}

In order to assess the error related to the batching of live point updates, independent nested samplings are performed (i) without any corrections, (ii) with only batch sorting, (iii) with a single iteration of the history correction, (iv) for two iterations of the history correction, and (v) for 10 iterations of the history correction for different batch sizes. The samplings are performed in a two-dimensional prior box with ranges $x_i \in [-7,7],\, i = 1,2$ and with a Gaussian likelihood:
\begin{equation} \label{eq:gaussian}
	\log\, \mathcal{L}(\vec{x}) = -\frac{1}{2} \sum_{i}^{\rm d} x_i^2\, ,
\end{equation}
where $\vec{x}=[x_1,x_2]$ and $d=2$. For a fixed number of live points, $N_{\rm live} = 10^3$, we compute evidences for $m/N_{\rm live} \in \{0.001,0.002,0.005,0.01,0.02,0.05,0.1,0.2,0.5,1\}$ using 50 independent nested samplings. The means and standard deviations of the resulting evidences for (i), (ii), and (iii) are shown in Figure~\ref{fig:cor1000} with the true value, $\log{\mathcal{Z}_{\rm true}} \approx\frac{d}{2}\log{\left(2\pi\right)} - d\log{(14)}$, subtracted.

From the figure it is clear that the correction does not matter too much for very low batch sizes. In particular, any differences for $m=1$ should only be due to noise since $m=1$ corresponds to regular sequential nested sampling. Notably, the naïve approach remains good until $m/N_{\rm live}$ of a few percent, but above that, the corrections start to become very important. Below $m/N_{\rm live}\sim10\%$, only doing batch sorting seems to be sufficient, while larger batch sizes require the history correction for the result not to diverge. For this simple example, we note that there is no significant improvement of using more iterations of history correction, so using a single iteration should be sufficient in similar cases. A good ``rule of thumb'' seems to be to keep $m/N_{\rm live}$ below $10\%$ and employ a single iteration of the history correction in order for the results to closely resemble what could be expected of sequential nested sampling. Additional tests (not shown) in higher dimensions ($d \gtrsim 10$) show that the history correction is not always able to keep the error from diverging close to $m/N_{\rm live} \rightarrow 1$ for a moderate number of correction iterations, but it is still able to keep the error small for $m/N_{\rm live} \lesssim 10\%$.

\section{Comparison with other nested samplers}\label{sec:comparison}

We will compare our nested sampler with two state-of-the-art public nested samplers, UltraNest and JAXNS. UltraNest features a sophisticated sampling algorithm with a variable number of live points. It is very efficient when likelihood calls dominate the run-time, and even though it runs exclusively on a CPU, it is able to use GPU compatible likelihood functions and utilise their vectorisation. JAXNS is a nested sampler implemented in the JAX framework which enables XLA acceleration and GPU compatibility as well. It is therefore static-memory in order for JIT compilation to be most efficient.

We will compare the three nested samplers on different analytic likelihood functions and dimensionalities, and compare their run-times and evidence results. Again, we will only compare performances on CPU, since these simple analytical functions run much faster on the CPU. Another reason is that UltraNest cannot run on a GPU (although it is compatible with likelihood functions running on a GPU) and JAXNS performs significantly worse in terms of run-time on a standard GPU because it enforces FP64 precision. Sampling on CPU, however, means that we cannot utilise the batching of live point updates efficiently, so all comparisons are with sequential nested sampling. In Section~\ref{sec:cosmo}, we will explore the benefits of sampling on a GPU when the likelihood functions are swapped by more complex and computationally expensive neural networks.

UltraNest and JAXNS are both run with default precision settings and for them we only vary the number of live points. Our sampler uses matching settings that closely resemble the defaults in JAXNS where the algorithmic differences are smallest. In particular, we adopt the same tolerance for the stopping criterion as well as the number ($5\times d$) of slice sampling iterations, but keep our sampler at FP32 precision throughout all tests. Since all likelihood functions used in these tests are analytic, their Bayesian evidences can be computed independently to high numerical precision and are used as the reference values below. When testing, the same analytic function is implemented in JAX for JAXNS and in \tf{} for our sampler and UltraNest. JAXNS JIT compiles the entire nested sampling, while our sampler only compiles a chunk of 1,000 iterations (hyperparameter) leading to a speed-up already on the initial run. UltraNest uses JIT compiled versions of the \tf{} likelihood functions. All reported run-times are for subsequent samplings after compilation and do therefore not include the compilation times. The compilation times are typically significantly larger for very fast samplings, but almost negligible for very slow samplings.

\subsection{Simple Gaussian}
\begin{figure}[tb]
	\centering
	\includegraphics[width=\textwidth]{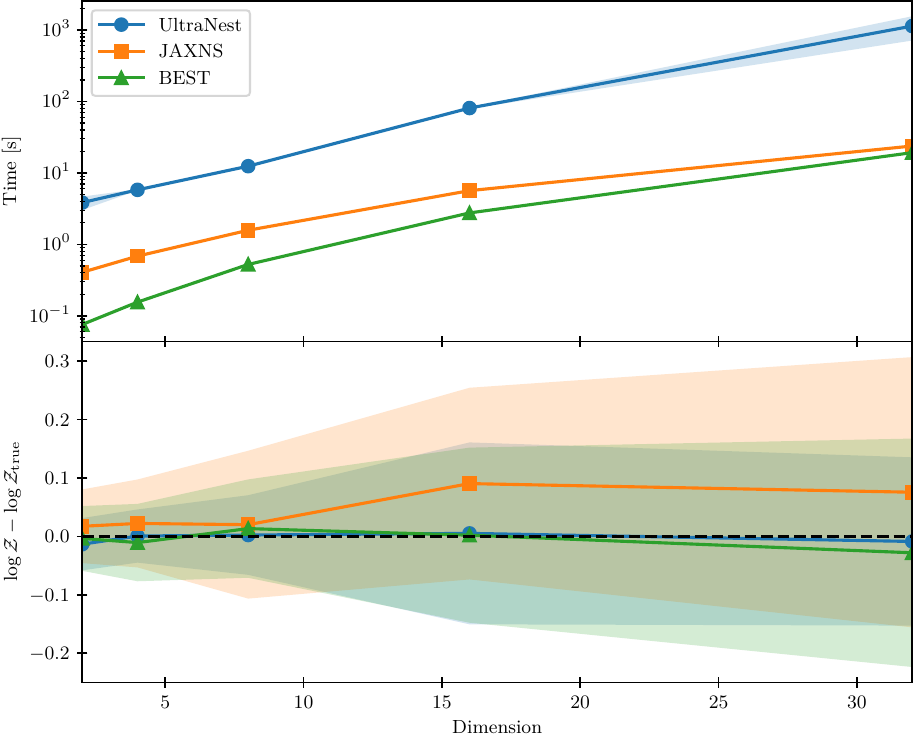}
	\caption{\textsl{Run-times (top) and errors on evidence (bottom) as a function of dimensionality of an uncorrelated Gaussian likelihood function for UltraNest, JAXNS, and our sampler in \textsc{best}. The results are averaged over 50 independent runs except for UltraNest only being averaged over 10 independent runs. The samplings were performed on a 14-core Apple M4 Pro CPU.}}
	\label{fig:compare_gauss}
\end{figure}

For this comparison, we reuse the simple uncorrelated Gaussian from equation~\eqref{eq:gaussian} with the same uniform prior box of $x_i \in [-7,7]$. We will however compare using dimensions of $d \in \{2,4,8,16,32\}$. Figure~\ref{fig:compare_gauss} shows the run-times and evidence deviations for all three nested samplers as a function of dimensionality. The results are averaged over 50 independent runs with 1,000 live points for our sampler and JAXNS, while we only use 10 independent runs for the UltraNest results due to run-time limitations. Due to the dynamic nature of UltraNest, we can only specify the minimum number of live points to 1,000.

UltraNest gives the most accurate evidence estimates in this test, but it is also much slower than the other two. Our sampler is noticeably faster than JAXNS in low dimensions, but the relative difference shrinks for higher dimensions. Our sampler is also more accurate than JAXNS across dimensions for this specific setup, but it should, however, be emphasised that even an error of 0.1 in $\log{\mathcal{Z}}$ is much less than what is significant for model comparisons according to Jeffreys' scale~\cite{Trotta:2008qt}.

\subsection{Rosenbrock function}
The Rosenbrock function~\cite{10.1093/comjnl/3.3.175} is regarded as a standard test case for optimisation routines. It features an almost-flat, narrow, parabolic valley in which the global minimum resides. Its negative will serve as the logarithm of the likelihood function within a uniform prior box of $x_i \in [-10, 10]$ for $i=1,2$. It is defined as:
\begin{equation}
	\log{\mathcal{L}}(\vec{x}) = -(a-x_1)^2 - b(x_2-x_1^2)^2\,,
\end{equation}
where we will use the standard values $a=1$ and $b=100$.

\begin{figure}[tb]
	\centering
	\includegraphics[width=\textwidth]{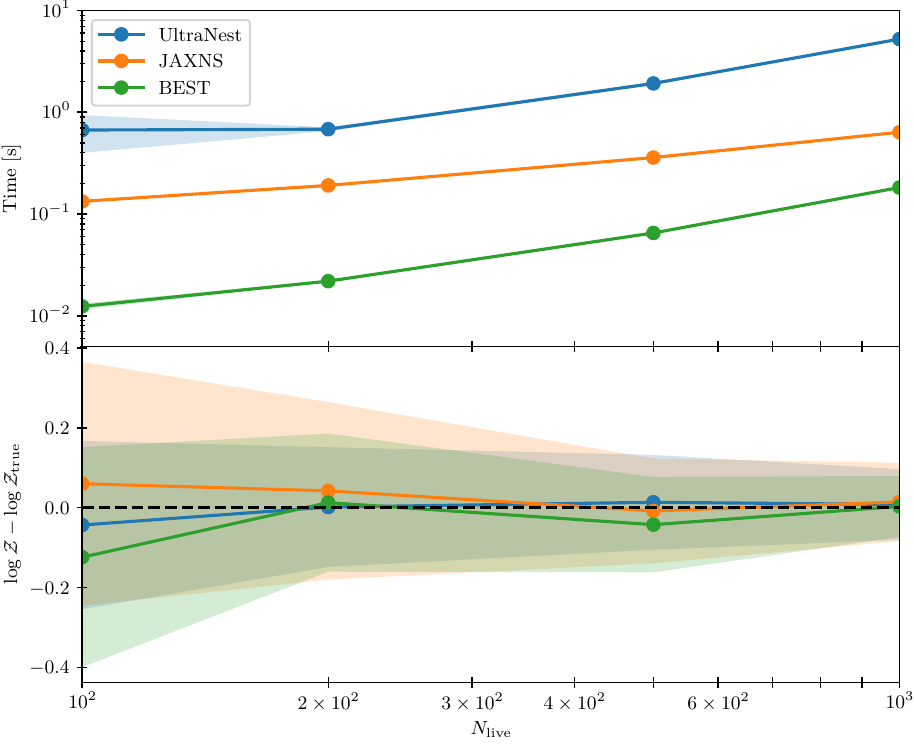}
	\caption{\textsl{Run-times (top) and errors on evidence (bottom) as a function of the number of live points for UltraNest, JAXNS, and our sampler in \textsc{best} when sampling the Rosenbrock function. The results are averaged over 100 independent runs except for UltraNest only being averaged over 10 independent runs. The samplings were performed on a 14-core Apple M4 Pro CPU.}}
	\label{fig:compare_rosenbrock}
\end{figure}

We test the three samplers using different numbers of live points (or minimum number of live points for UltraNest), i.e., $N_{\rm live} \in \{100,200,500,1,000\}$, and this time, we average over 100 independent runs for our sampler and JAXNS, while only using 10 independent runs for UltraNest. The run-times and errors on the evidence are shown in Figure~\ref{fig:compare_rosenbrock} as a function of the number of live points. Again, our sampler outperforms the other two in terms of run-time, but as the number of live points grows, the difference with the run-time of JAXNS seems again to shrink.

UltraNest again gives the most accurate evidence estimates in this test, but all three samplers seem to converge nicely with more live points. It should again be emphasised that UltraNest is not restricted to a fixed number of live points, so the very good results even with only 100 live points (as a minimum) can partly be attributed to the dynamic nature of the sampler.

\subsection{Multimodal distribution}
\begin{figure}[tb]
	\centering
	\includegraphics[width=\textwidth]{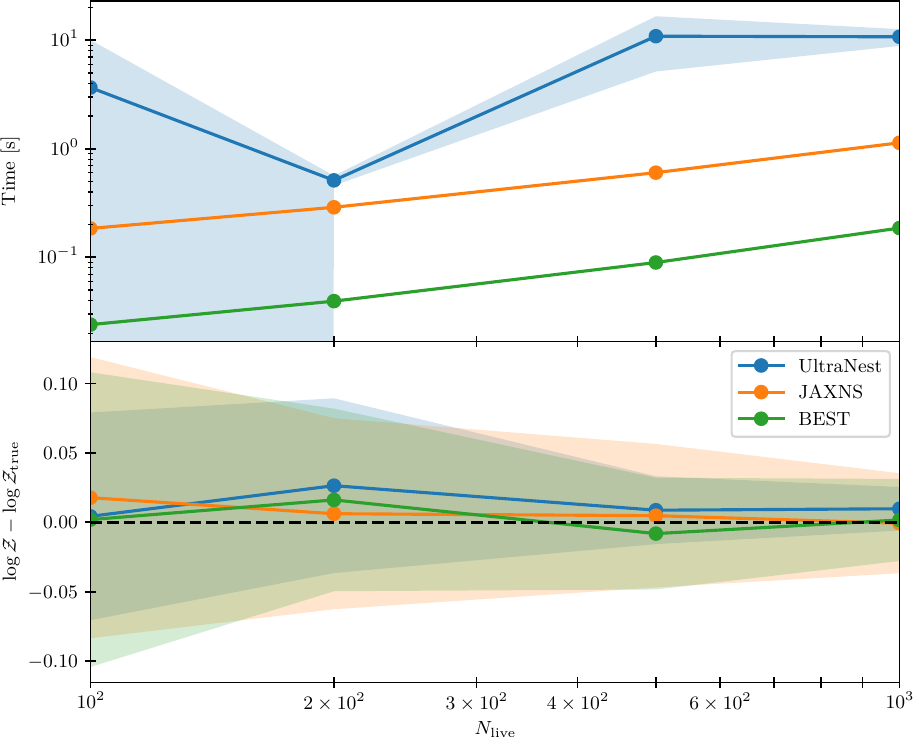}
	\caption{\textsl{Run-times (top) and errors on evidence (bottom) as a function of the number of live points for UltraNest, JAXNS, and our sampler in \textsc{best} when sampling a 2D array of 49 disjoint Gaussian modes. The results are averaged over 100 independent runs except for UltraNest only being averaged over 10 independent runs. The samplings were performed on a 14-core Apple M4 Pro CPU.}}
	\label{fig:compare_array}
\end{figure}

Nested samplers are usually expected to perform well on multimodal distributions, and to test this, we will employ the same array of two-dimensional Gaussians described by equation~\eqref{eq:array} in the same $[-20,20]^2$ prior box. We again average over 100 independent runs for our sampler and JAXNS, while only averaging over 10 independent runs for UltraNest. The run-times and errors on the evidence are shown in Figure~\ref{fig:compare_array} as a function of live points. The trend is generally the same as for the other tests, i.e., the evidence results converge with more live points for all samplers.

For this problem, UltraNest shows somewhat larger evidence deviations in this test, but the errors are still well below what would be considered statistically significant. With too few live points, the run-time of UltraNest is very noisy and this might be due to the difference in the likelihood-constrained sampling in the three samplers. UltraNest by default uses MLFriends~\cite{Buchner:2021kpm} to explore multiple modes instead of slice sampling and with a limited number of live points, this might be more difficult in this example with an array of many disjoint modes as opposed to the other examples involving only a single mode. The spread is still, however, quite small when the number of live points is increased and UltraNest thus remains a very accurate nested sampler, though with the drawback of requiring substantially more compute time.

\begin{table}[]
\centering
\begin{tabular}{l|cc}
                                   & \textbf{Lower limit} & \textbf{Upper limit} \\ \hline
\texttt{omega\_b}                  & 0       & 3       \\
\texttt{omega\_cdm}                & 0       & 0.5     \\
\texttt{100*theta\_s}              & 0       & 2       \\
\texttt{ln10\^\{10\}A\_s}          & 0       & 5       \\
\texttt{n\_s}                      & 0       & 2       \\
\texttt{tau\_reio}                 & 0.004   & 1       \\
\texttt{A\_cib\_217}               & 0       & 200     \\
\texttt{xi\_sz\_cib}               & 0       & 1       \\
\texttt{A\_sz}                     & 0       & 10      \\
\texttt{ps\_A\_100\_100}           & 0       & 400     \\
\texttt{ps\_A\_143\_143}           & 0       & 400     \\
\texttt{ps\_A\_143\_217}           & 0       & 400     \\
\texttt{ps\_A\_217\_217}           & 0       & 400     \\
\texttt{ksz\_norm}                 & 0       & 10      \\
\texttt{gal545\_A\_100}            & 0       & 50      \\
\texttt{gal545\_A\_143}            & 0       & 50      \\
\texttt{gal545\_A\_143\_217}       & 0       & 100     \\
\texttt{gal545\_A\_217}            & 0       & 400     \\
\texttt{galf\_TE\_A\_100}          & 0       & 10      \\
\texttt{galf\_TE\_A\_100\_143}     & 0       & 10      \\
\texttt{galf\_TE\_A\_100\_217}     & 0       & 10      \\
\texttt{galf\_TE\_A\_143}          & 0       & 10      \\
\texttt{galf\_TE\_A\_143\_217}     & 0       & 10      \\
\texttt{galf\_TE\_A\_217}          & 0       & 10      \\
\texttt{calib\_100T}               & 0       & 3000    \\
\texttt{calib\_217T}               & 0       & 3000    \\
\texttt{A\_planck}                 & 0.9     & 1.1
\end{tabular}
\caption{\textsl{Lower and upper bounds for the model parameters in the neural network emulator. These bounds are used as the prior ranges for the nested sampling.}}
\label{tab:prior}
\end{table}

\section{Using with cosmological emulators}\label{sec:cosmo}
The entire motivation behind implementing a pure \tf{} nested sampler in \textsc{best} was to efficiently utilise cosmological likelihood emulators from CLiENT. These neural network emulators are much more demanding than simple analytic functions as used previously in the paper, and even though they are orders of magnitude faster than running a conventional likelihood pipeline, we require GPUs to fully take advantage of the batching implemented in our nested sampler.

Using an emulator from Ref.~\cite{Janken:2025wlq}, we obtain the Bayesian evidence for the $\Lambda$CDM model with the following data sets:
\begin{itemize}
	\item Planck 2018 high-$\ell$ TTTEEE, low-$\ell$ TT+EE, and lensing~\cite{Planck:2018vyg,Planck:2019nip}.
	\item Baryon Acoustic Oscillations (BAO) measurements from BOSS DR12~\cite{boss2016}, the main galaxy sample of BOSS DR7~\cite{ross2014} and 6dFGS~\cite{Beutler:2011hx}.
\end{itemize}

\begin{figure}[tb]
	\centering
	\includegraphics[width=\textwidth]{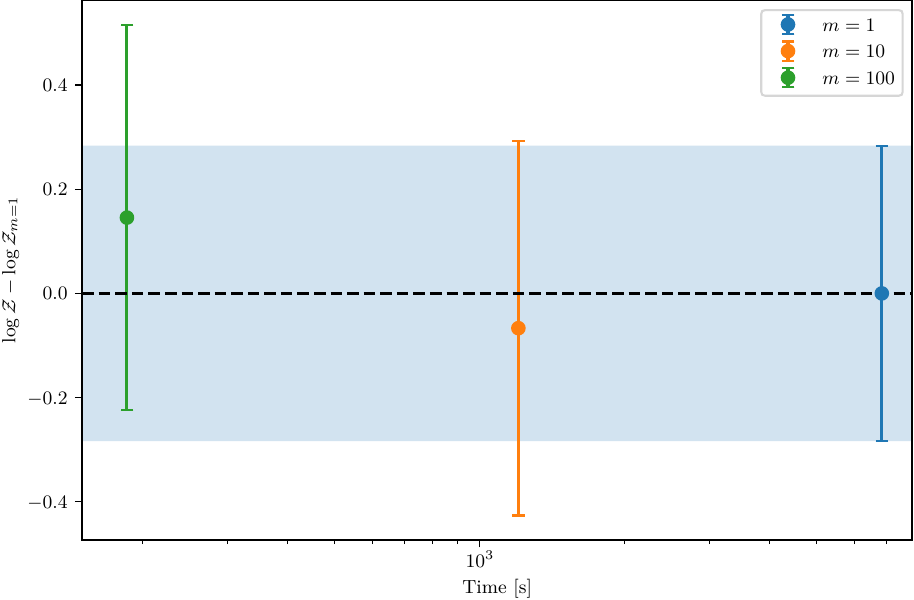}
	\caption{\textsl{Run-times and evidence estimates for nested samplings of the cosmological likelihood emulator using 1,000 live points and different batch sizes, $m$. The sequential ($m=1$) result was averaged over three independent samplings whereas the samplings with larger batch sizes were averaged over 10 independent runs. The blue band represents the $m=1$ result as a benchmark.}}
	\label{fig:nn}
\end{figure}

This is a 27-dimensional parameter space including the 21 Planck nuisance parameters, and the prior box used in the sampling is reported in Table~\ref{tab:prior}. A nested sampling with 1,000 live points and without batched updates ($m=1$) serves as our benchmark for this test, but since a sampling such as this one takes several hours (even on a GPU) we only use 7 independent samplings and report the evidence result and error as the mean and standard deviation of their individual evidence results, respectively. Figure~\ref{fig:nn} shows the evidence estimates and run-times of samplings using different batch sizes, $m$. All points with $m>1$ are averaged over 10 independent samplings with the standard deviation of the individual evidence results as the error. The samplings with different batch sizes seem to agree well within the reported errors, and the discrepancy is well below anything significant for model comparisons according to Jeffreys' scale. Including the batching of updates, however, decreases the overall run-time immensely, so a larger batch size (still around or below 10\% of $N_{\rm live}$ though) can therefore be preferable when comparing models expected to yield a strong or medium evidence for one over the other according to Jeffreys' scale. If two models turn out to have very similar evidence results, one can always decrease the batch size and/or run a sequential sampling afterwards to solidify the result. Due to the limited number of independent runs, this test serves as a demonstration of consistency and performance rather than a high-precision measurement of the batching bias.




\section{Conclusion}\label{sec:conclusion}
In this paper, we have presented a new nested sampler implemented in \textsc{best}, extending the framework beyond its existing MCMC and profile-likelihood capabilities. The sampler is written entirely in TensorFlow and designed to efficiently exploit XLA compilation, vectorisation, and GPU acceleration. Its main algorithmic addition is a corrected scheme for updating several live points simultaneously.

We have shown that this batching introduces an additional source of error in the evidence estimate, but that the effect can be kept small using the sorting and history corrections introduced here. In particular, keeping the batch size below roughly $10\%$ of the number of live points provides a good compromise between accuracy and parallel efficiency. The sampler was tested on Gaussian, Rosenbrock, and multimodal likelihoods and compared to JAXNS and UltraNest, recovering consistent evidence estimates while generally reducing the run-time. Clustering was found to be useful mainly for sufficiently complex or expensive likelihoods, where the reduction in likelihood evaluations outweighs the added clustering overhead.

Finally, using a 27-dimensional cosmological likelihood emulator, we showed that batching can substantially reduce the wall-clock time while maintaining evidence estimates consistent with sequential sampling within the reported uncertainties. This makes the new nested sampler particularly useful in the regime for which \textsc{best} is intended: fast, vectorised likelihoods and emulators where the sampling algorithm itself can otherwise become a significant part of the total computational cost.

\section*{Acknowledgements}
We acknowledge computing resources from the Centre for Scientific Computing Aarhus (CSCAA). The work presented here is supported by the Carlsberg Foundation, grant CF24-1944.


\bibliographystyle{utcaps}
\bibliography{NS2026}

\end{document}